%% file: main.tex
\documentclass[runningheads]{llncs}

\usepackage{eccv}

\usepackage{eccvabbrv}

\usepackage{graphicx}
\usepackage{booktabs}

\usepackage[accsupp]{axessibility}  

\usepackage{hyperref}
\newcommand*\circled[1]{\tikz[baseline=(char.base)]{
            \node[shape=circle,draw,inner sep=1pt] (char) {#1};}}
\usepackage{orcidlink}
\usepackage{graphicx}
\usepackage{amsmath}
\usepackage{amssymb}
\usepackage{booktabs}
\usepackage{multirow}
\usepackage{float}
\usepackage{todonotes}
\usepackage{float}
\usepackage{wrapfig}

\begin{document}

\title{Rawformer: Unpaired Raw-to-Raw Translation for Learnable Camera ISPs} 

\titlerunning{Rawformer}

\author{Georgy Perevozchikov\inst{1}\orcidlink{0009-0009-7176-6242} \and
Nancy Mehta\inst{1}\orcidlink{0000-0002-1249-8577} \thanks{Corresponding Author}\and
Mahmoud Afifi\inst{2}\orcidlink{0000-0003-0125-4945}\thanks{Now at Google} \and Radu Timofte\inst{1}\orcidlink{0000-0002-1478-0402}}

\authorrunning{G.~Perevozchikov et al.}

\institute{Computer Vision Lab, CAIDAS \& IFI, University of W\"urzburg,\\ John Skilton Str. 4a, 97074 W\"urzburg, Germany
\email{\{georgii.perevozchikov,nancy.mehta,radu.timofte\}@uni-wuerzburg.de} \and
York University, 4700 Keele St, Toronto, Ontario, Canada, M3J 1P3\\
\email{m.3afifi@gmail.com}
}

\maketitle

\input{sections_cr/abstract}
\input{sections_cr/intro}
\input{sections_cr/related_work}
\input{sections_cr/proposed_method}
\input{sections_cr/experiments}

\input{sections_cr/conclusion}

\section*{Acknowledgments}
This work was partly supported by The Alexander von Humboldt Foundation.

%
%

\title{Supplemental Materials of\\Rawformer: Unpaired Raw-to-Raw Translation for Learnable Camera ISPs} 

\titlerunning{Rawformer (Supplemental Materials)}

\author{Georgy Perevozchikov\inst{1}\orcidlink{0009-0009-7176-6242} \and
Nancy Mehta\inst{1}\orcidlink{0000-0002-1249-8577}\thanks{Corresponding Author} \and
Mahmoud Afifi\inst{2}\orcidlink{0000-0003-0125-4945}\thanks{Now at Google} \and Radu Timofte\inst{1}\orcidlink{0000-0002-1478-0402}}

\authorrunning{G.~Perevozchikov et al.}


\institute{Computer Vision Lab, CAIDAS \& IFI, University of W\"urzburg,\\ John Skilton Str. 4a, 97074 W\"urzburg, Germany
\email{\{georgii.perevozchikov,nancy.mehta,radu.timofte\}@uni-wuerzburg.de} \and
York University, 4700 Keele St, Toronto, Ontario, Canada, M3J 1P3\\
\email{m.3afifi@gmail.com}
}


\maketitle

\input{sections_cr/supp_materials}

%
%

\bibliographystyle{splncs04}
\bibliography{egbib}
\end{document}

%% file: sections_cr/abstract.tex
\begin{abstract}
 
Modern smartphone camera quality heavily relies on the image signal processor (ISP) to enhance captured raw images, utilizing carefully designed modules to produce final output images encoded in a standard color space (e.g., sRGB). Neural-based end-to-end learnable ISPs offer promising advancements, potentially replacing traditional ISPs with their ability to adapt without requiring extensive tuning for each new camera model, as is often the case for nearly every module in traditional ISPs. However, the key challenge with the recent learning-based ISPs is the urge to collect large paired datasets for each distinct camera model due to the influence of intrinsic camera characteristics on the formation of input raw images. This paper tackles this challenge by introducing a novel method for unpaired learning of raw-to-raw translation across diverse cameras. Specifically, we propose Rawformer, an unsupervised Transformer-based encoder-decoder method for raw-to-raw translation. It accurately maps raw images captured by a certain camera to the target camera, facilitating the generalization of learnable ISPs to new unseen cameras. Our method demonstrates superior performance on real camera datasets, achieving higher accuracy compared to previous state-of-the-art techniques, and preserving a more robust correlation between the original and translated raw images. The codes and the pretrained models are available at \url{https://github.com/gosha20777/rawformer}.
\end{abstract}

%% file: sections_cr/intro.tex
\section{Introduction}
\label{sec:intro}

\begin{figure}[t]
  \centering
   \includegraphics[width=0.95\linewidth]{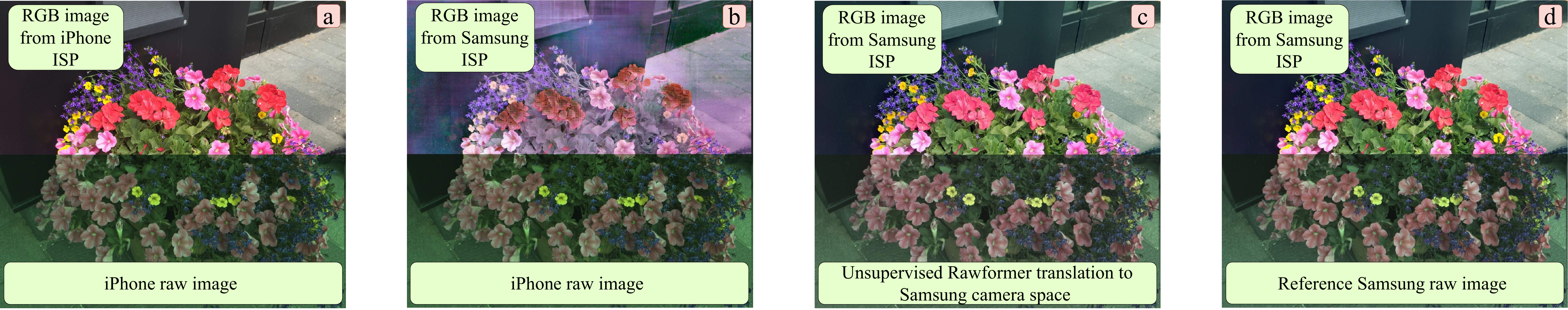}

   \caption{\textbf{We introduce Rawformer}, an unsupervised method for raw-to-raw translation that allows the utilization of pre-trained neural-based ISPs to process raw images captured by previously unseen cameras.  Shown are raw and sRGB images processed by a neural-based ISP \cite{wirzberger2022lan}. a) Raw image from an iPhone X rendered by the neural-based ISP trained on iPhone X's raw images. b) Raw image from an iPhone X rendered by the neural-based ISP trained on Samsung S9's raw images. c) iPhone X raw image translated to Samsung S9's raw space using our method, then processed by the Samsung S9 neural-based ISP. d) Raw image captured by Samsung S9's camera rendered by its native camera ISP, provided as a reference for visual comparison.}
   \label{fig:visualisation}
\end{figure}

Unlike the high-quality digital single-lens reflex (DSLR) cameras, mobile phone cameras possesses inherent limitations owing to their smaller sensor size and fixed lenses with limited optical capabilities \cite{wronski2019handheld}. The mobile camera image signal processor (ISP) aims to overcome these limitations by applying a series of carefully designed modules (e.g., \cite{vsindelavr2013image, barron2017fast, herrmann2020learning, a2021beyond, jiang2022fast, conde2023perceptual}) to enhance the quality of the final image rendered in a standard color space, such as sRGB. These modules work in tandem to optimize key parameters, such as sharpness, color accuracy, and noise reduction, resulting in a visually pleasing sRGB image from the raw image captured by the camera \cite{hasinoff2016burst, delbracio2021mobile}.

Despite yielding promising results, achieving compatibility among camera ISP modules for each new camera device requires extensive tuning and adjustment efforts to attain the desired image quality. This motivates the use of deep neural networks to replace individual camera ISP modules with a single neural-based unit capable of accomplishing the task \cite{schwartz2018deepisp,  dai2020awnet, Ignatov_2021_CVPR}. 
Nonetheless, the training of such neural-based ISPs demands an extensive dataset consisting of raw images taken with a specific camera, accompanied by corresponding ``ground-truth'' sRGB images (typically generated by a high-quality DSLR camera) \cite{ignatov2020replacing, Ignatov_2021_CVPR}.

An issue emerges with the introduction of a new camera model, as pre-trained neural-based ISPs might encounter difficulty in accurately rendering raw images captured by the new camera (refer to Fig. \ref{fig:visualisation}(b)). This difficulty arises from the potential distinct characteristics of the new camera, such as sensor sensitivities, which influence the formation of its raw images \cite{afifi2019sensor, afifi2021semi}. Consequently, inconsistencies arise in the interpretation of raw RGB colors between cameras \cite{afifi2021semi}, hindering the generalization of neural-based ISPs to new cameras not encountered during training. To address this issue, re-training or fine-tuning on a new paired dataset becomes necessary, in which the raw images are sourced from the target new camera. This process mirrors similar challenges encountered in traditional ISP development for new cameras.

An alternative strategy to address this challenge is domain adaptation, where the raw images from the new target camera are mapped to the raw space of the training camera to emulate those captured by the original camera used for training. In light of this, \cite{afifi2021semi} proposed a semi-supervised learning framework for raw image translation, but the approach outlined in \cite{afifi2021semi} inherits the drawbacks associated with the CNN-based architecture, notably the limited receptive field and inefficient encoding of global information. Thus, we present a fully unsupervised Transformer-based raw-to-raw method, dubbed Rawformer, that maps raw images between camera models by efficiently encoding their global and semantic correlations. Our method eliminates the need for a paired raw-sRGB dataset from each new target camera, enabling the utilization of a neural-based ISP trained on a specific camera's raw images to process raw images taken by new cameras with different characteristics without necessitating any re-training; see Fig. \ref{fig:visualisation}(c). Our contributions are summarized as follows:
\begin{itemize}

\item We propose Rawformer, a \textit{fully unsupervised} encoder-decoder Transformer-based method designed for raw-to-raw translation facilitating the reuse of neural-based ISPs. Rawformer achieves state-of-the-art results in raw-to-raw translation, leading to better generalization for cross-camera learnable ISP rendering.

\item We introduce contextual-scale aware downsampler and upsampler blocks that efficiently summarize the local-global contextual details in mixed scale representations via its condensed query attention block and scale perceptive feed-forward network.

\item A novel cross-domain attention-driven discriminator is proposed along with a specialized discriminator head for stabilizing the network training.

\end{itemize}

%% file: sections_cr/related_work.tex
\section{Related Work}

\subsection{Neural-Based ISP} 

Several neural-based ISPs methods have been introduced \cite{chen2018learning, zamir2020cycleisp, xing2021invertible, Ignatov_2021_CVPR, zhang2021learning, jeong2022rawtobit, he2024enhancing}, with the majority relying on CNN-based U-Net-like architecture~\cite{ronneberger2015u}. These include earlier approaches, such as DeepISP~\cite{schwartz2018deepisp} and PyNET~\cite{ignatov2020replacing}, as well as more recent methods, such as MW-ISPNet~\cite{ignatov2020aim}, AW-Net~\cite{dai2020awnet}, and LAN~\cite{wirzberger2022lan}, which achieve improved image restoration quality by integrating discrete wavelet transform (DWT) and double attention modules (DAM) techniques. Such learnable ISPs are typically trained to map raw images captured by a \textit{specific camera} model to target high-quality sRGB ``ground-truth'' images produced by some target camera ISP (usually with higher quality than the mobile native ISP; e.g., a DSLR camera \cite{ignatov2020replacing, Ignatov_2021_CVPR}). A significant challenge for neural-based ISPs is the need for paired datasets of raw-sRGB images. Collecting such datasets is tedious and typically involves using specific equipment. Additionally, it may require post-processing techniques to align the raw images with the target sRGB images \cite{ignatov2020replacing}. The challenge intensifies when a mobile phone manufacturer introduces a new camera model with different characteristics, which happens frequently \cite{tominaga2021measurement}, requiring the recollection of paired raw-sRGB datasets for each new camera model. 

One solution is to render images to the sRGB space from the device-independent CIE XYZ space \cite{afifi2021cie}. However, this requires calibration for each camera model, and accuracy depends on specific camera conditions (e.g., the Luther condition) \cite{finlayson2020designing}. In contrast, our work introduces an unsupervised raw-to-raw mapping technique that eliminates the need for paired raw-sRGB datasets for re-training or camera calibration. Our method maps raw images from new cameras to the camera raw space utilized during the training of neural-based ISPs, enabling the reuse of pre-trained ISPs.

\subsection{Image-to-Image Domain Adaptation}
Numerous generative adversarial network (GAN)-based methods address unpaired image-to-image (I2I) translation challenges~\cite{pang2021image}. Notably, CycleGAN~\cite{zhu2017unpaired}, DualGAN~\cite{yi2017dualgan}, UNIT~\cite{liu2017unsupervised}, STARGAN~\cite{choi2018stargan}, SEAN~\cite{zhu2020sean}, and U-GAT-IT~\cite{kim2019u} focus on cycle consistency, employing dual generator networks guided by a cycle-consistency constraint. ACLGAN~\cite{zhao2020unpaired} and CUT~\cite{park2020contrastive} represent advancements in this field. ACLGAN relaxes the cycle consistency constraint, favoring an adversarial constraint, potentially enhancing translation quality. CUT utilizes a contrastive loss, eliminating the need for multiple generators and enabling faster training. ITTR further refines CUT's performance by altering the generator architecture. CycleGAN-based model, such as UVCGAN~\cite{torbunov2023uvcgan} incorporate U-Net-like generators with vision Transformers (ViTs).

In computational photography, attempts have been made for domain adaptation, including synthesizing raw night images from raw daylight images to enhance neural-ISP rendering quality~\cite{punnappurath2022day}, few-shot domain adaptation for low-light image enhancement~\cite{prabhakar2023few}, and mapping graphics images to a target camera's raw space~\cite{seo2023graphics2raw}. Raw-to-raw translation has received even limited attention. Classical approaches require a calibration object to learn global mapping \cite{nguyen2014raw}, while recent work~\cite{afifi2021semi} proposes a semi-supervised CNN learning-based method using a small set of perfectly aligned images. Despite being effective at exploiting the local semantics, this CNN-based method lacks the potential to exploit the long-range/global semantics between different domains. In contrast, our method aims at solving the raw mapping by exploiting both local and global contextual information through \textit{fully} unsupervised training, achieving state-of-the-art results.

%% file: sections_cr/proposed_method.tex
\section{Proposed Method}
\begin{figure}[t]
  \centering
  \includegraphics[width=1.\linewidth]{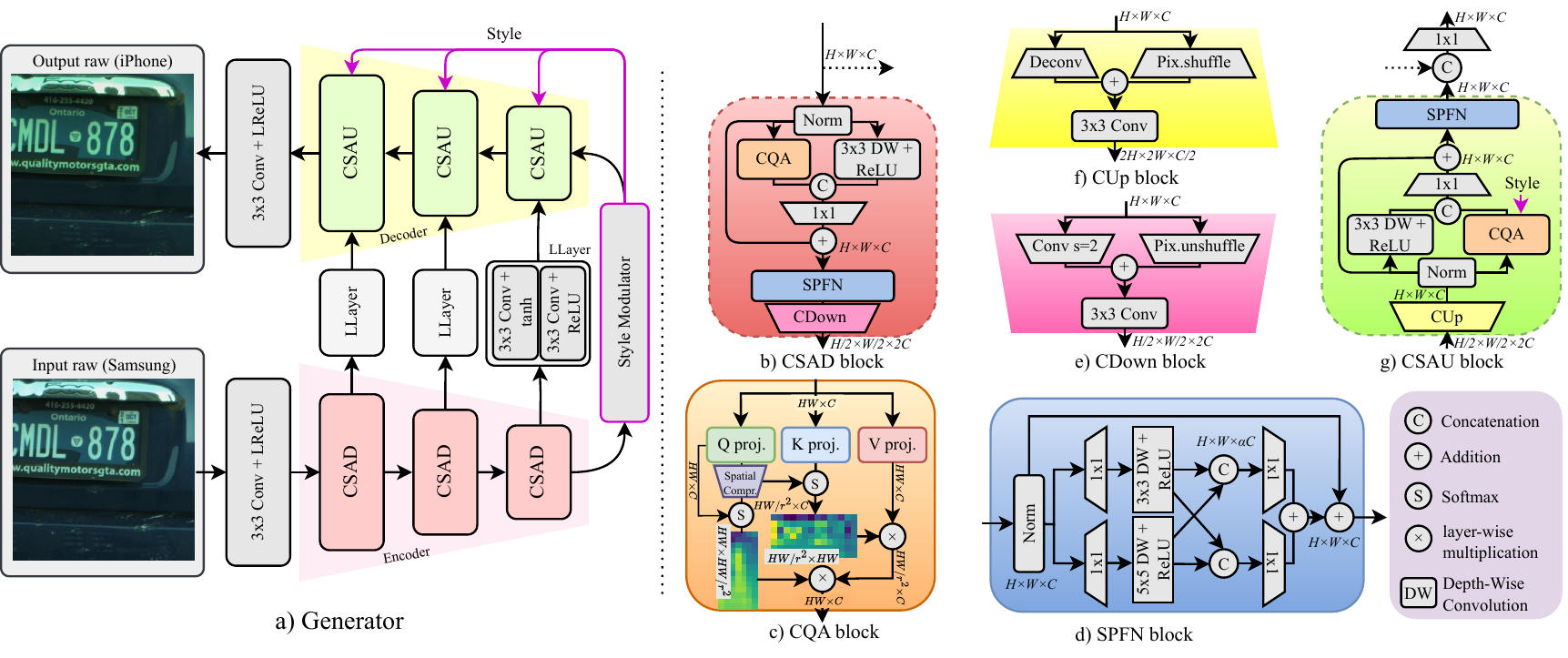}
  \caption{ a) \textbf{Overview of the generator architecture (a) of Rawformer.} The primary components of the generator: b) contextual-scale aware downsampler block (CSAD), c) condensed query attention block (CQA), d) scale perceptive feed-forward network (SPFN), e) composite downsampling block (CDown), f) composite upsampler (CUp) block, and g) contextual-scale aware upsampler block (CSAU).}
  \label{fig:generator}
\end{figure}

Existing raw-to-raw methodologies either rely on supervised approaches requiring paired datasets e.g., \cite{nguyen2014raw}) or semi-supervised approaches (e.g., \cite{afifi2021semi}) with limited accuracy. Additionally, domain adaptation techniques  (e.g., \cite{park2020contrastive, zhu2017unpaired}), which aim to extract local semantics between source and target domains, often fail to capture overall semantics and context-dependent relationships. Our proposed method, Rawformer, addresses these limitations by processing raw images in a \textit{fully} unsupervised manner. Utilizing an encoder-decoder architecture within a CycleGAN framework, Rawformer effectively captures global semantics and maintains intrinsic consistency between translated images. The generator (Fig. \ref{fig:generator}(a)) in Rawformer employs contextual-scale aware downsampler and upsampler blocks to enhance content rendering and structural integrity. The discriminator (Fig. \ref{fig:discriminator}(a))  stabilizes training by incorporating key discriminator features, thereby preventing model collapse and reducing redundant computations. In the following, we provide a comprehensive elucidation of the proposed components.

\subsection{Generator's Encoder Network}
As shown in Fig. \ref{fig:generator}(a), the encoder of the proposed generator in Rawformer comprises stacked contextual-scale aware downsampler (CSAD) blocks to bridge the semantic gap between the source and target domains. Each CSAD block, depicted in Fig. \ref{fig:generator}(b), offers a holistic approach of understanding the contextual dependencies efficiently between the domains by capturing both the global and local information via its condensed query attention (CQA) block, in parallel with a 3$\times$3 depth-wise convolution layer. It further handles the intrinsic correlations among the extracted features in a multi-scale manner via maintaining a synergistic collaboration between the scale perceptive feed-forward network (SPFN), and the composite downsampler (CDown) block. 

\subsubsection{Condensed Query Attention Block:}
Self-attention is highly effective in modeling long-range contextual dependencies by comparing each token with all others, facilitating seamless information propagation within the network \cite{zamir2022restormer}. However, its computational demands pose challenges for tasks like unpaired raw-to-raw translation, where maintaining both visual quality and semantic consistency is crucial.  To address these concerns and overcome the constraints of computation, we introduce the condensed query attention (CQA) block; see Fig. \ref{fig:generator}(c). Here, we make two key modifications to self-attention to enhance learning efficiency. Firstly, we spatially compress the query vector using average pooling followed by linear projection, and downsample the resulting feature map by a factor of $r$ to confine the crucial information. This results in condensed queries,$(\textbf{Q}_c)$ used for similarity comparison, as shown in Eq. \ref{eq:QM1}. Secondly, to balance global and regional modeling with computational efficiency, attention is applied horizontally and vertically to eliminate redundant computations. The CQA block operation is defined as:
\begin{equation}
\begin{array}{l}
    \,\,\,\,\,\,\,\,\,\,\,\,\,\,\,\,\, CQA(\textbf{Q}_c, \textbf{K},  \textbf{V}) = \textbf{A}_U.(\textbf{A}_H\cdot\textbf{V} ) ;\,\,\,\,\,\, \textbf{Q}_c = W_p(AvgPool(\textbf{Q}))\\
      \,\,\,\,\,\,\,\,\,\,\,\,\,\,\,\,\,  \textbf{A}_H = Softmax( \frac{\textbf{Q}_c\textbf{K}^T}{\sqrt{d}} ) ,\,\,\,\, \textbf{A}_U = Softmax( \frac{\textbf{Q} \textbf{Q}_c^T}{\sqrt{d}} )  \\
      \\
\end{array}
\label{eq:QM1}
\end{equation}
\noindent where, the duet of key ($\textbf{K}$), and value $(\textbf{V})$ vectors are obtained through plain linear convolutions, and the dimensions of each of these features are shown in the Fig. \ref{fig:generator}(c). $W_{p}{(.)}$ denotes a 1$\times$1 pointwise convolution, and $\textbf{A}_H$, and $\textbf{A}_U$ represents the compressed attention map between the key-condensed query pair horizontally and query-condensed query pair (for focusing on fine grained contextual information) vertically, respectively. \textit{The choice of operations and the reason for condensing the queries is investigated in the supplementary of the paper.}

\subsubsection{Scale Perceptive Feed-forward Network:}

To proficiently capture the intricacies of the local image structure across various scales by encoding the information from neighboring pixel locations, we propose a scale perceptive feed-forward network (SPFN); see Fig. \ref{fig:generator}(d). Unlike the conventional feed-forward networks that integrate single-scale depth-wise convolutions \cite{zamir2022restormer}, we emphasize on enhancing the locality and inter-correlations among the incoming raw features of different domains by inserting multi-scale depth-wise convolutions. Given an input tensor at layer $s$ and scale $i$, such that ${\textbf{X}}_{s}^i \in {\mathbb{R}^{ H^i \times W^i \times C }}$, SPFN is formulated as:
\begin{equation}
\begin{array}{l}
\textbf{X}_{s}^{1} = \phi(W_{d}^3(W_{p}^1(LN(\textbf{X}_{s-1}))));  \textbf{X}_{s}^{2} = \phi(W_{d}^5(W_{p}^2(LN(\textbf{X}_{s-1})))), \\
\,\,\,\,\,\,\,\,\,\,\,\,\,\,\,\,\,\,\,\,\,\,\,\,\,\,\,\,\,\,\,\,\,\,\  \textbf{Y}_{s} = W_{p}^3(\textbf{X}_{s}^1\circled{C}\textbf{X}_{s}^2)+\textbf{X}_{s-1}.
\end{array}
\label{eq:QM3}
\end{equation}
Here, $\phi$ represents a LeakyReLU activation function \cite{maas2013rectifier}, $W_{d}^k$ represents a depth-wise convolution with filter size $k$,   $LN$ denotes the layer normalization \cite{ba2016layer}, and $\circled{C}$ represents the concatenation operation. This enriches a more nuanced understanding of the underlying image features at different hierarchical levels.

\subsubsection{Composite Downsampler:}
In general, the essence of downsampling is the retrieval of the multi-scale information in the image. Unlike the common encoder architectures for various image signal processing pipelines \cite{liang2021cameranet,souza2022crispnet} that deploy strided convolutions for downsampling, we design a hybrid strategy of leveraging the benefits of information exchange between different resolution features. To adaptively recover the underlying content information from the input domain at multiple scales, we design a composite down (CDown) block that intelligently combines pixel-unshuffle (to prevent the information loss) and strided convolution (to enhance the model's expressiveness) as shown in Fig.~\ref{fig:generator}(e).  

\subsubsection{Style Modulator:}
Different cameras have their own distinctive characteristics to be handled or restored. To augment the expressiveness of Rawformer for handling camera perturbations, we design a style modulator in the generator using extended pixel-wise ViT\cite{torbunov2023uvcgan2}. It aids in calibrating the decoder features and subsequently encourages the recovery of crucial details while translation. As demonstrated in Fig. \ref{fig:generator}(a), the style modulator applies style tokens in each contextual aware upsampler (CSAU) block, where every style token models the weight of the query vector of the CQA block to enrich the style specific context. This addition of style token at each stage in the decoder helps in the flexible adjustment of the feature maps. Unlike StyleGANv2 \cite{karras2020analyzing}, where the style vectors are generated from the random prior, our approach resorts to learnable style token for conjecturing the important target style directly from the source domain, thus enhancing the consistency and control in raw image translation. \textit{More details on style modulation are provided in the supplementary.}

\subsection{Generator's Decoder Network}
 Unlike the popular additive or multiplicative skip connections between the encoder-decoder, we also present a learnable linking layer (LLayer) as shown in the Fig.~\ref{fig:generator}(a) to facilitate the feature preservation in the contextual scale aware upsampler (CSAU) block. The overall design of the CSAU block closely resembles that of the CSAD block, ensuring consistency and coherence in the network architecture in a bottom-up fashion. The primary distinction lies within the composite upsampling block, where for inducing the relevant spatial information into the target domain for the incoming features from the style modulator, we design a composite upsampler block (CUp). It embodies a parallel combination of pixel-shuffle (for preventing the checkerboard artifacts in the target domain) and deconvolution (for combating with aliasing) as shown in Fig.~\ref{fig:generator}(f) and (g).

\begin{figure}
   \centering
   \includegraphics[width=0.85\linewidth]{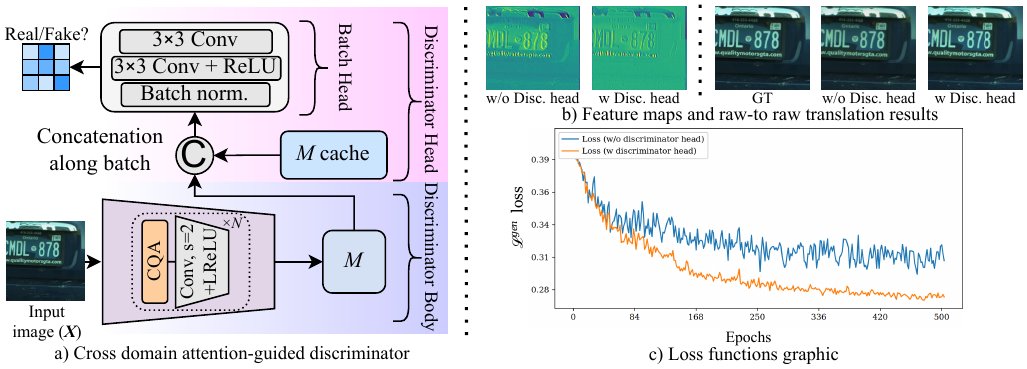}
   \caption{a) The proposed cross-domain attention-guided discriminator. b) Feature map visualizations without (w/o) and with (w) the discriminator head. The inclusion of the discriminator head aids in refining the overall results, and the discriminator training (as shown in c) benefits from the inclusion of the discriminator head, particularly with the cache ($M$) component.}
   \label{fig:discriminator}
\end{figure}

\subsection{Cross-Domain Attention-Guided Discriminator}

Leveraging batch statistics is an effective strategy to mitigate mode collapse and enhance diversity in GAN models~\cite{salimans2016improved}. Traditional minibatch discrimination techniques require large batch sizes~\cite{karras2017progressive}, posing challenges due to GPU limitations. Inspired by \cite{torbunov2023uvcgan}, we address this by decoupling batch size from the minibatch discrimination approach. Our core innovation involves using a cache of previous discriminator features as an alternative to larger batch sizes. During training, batch features are stored in four separate caches $(M)$; real images from the source and target cameras, and fake images from both domains. This memory bank-style cache utilizes a composite discriminator with a head and body. The head captures batch statistics via batch normalization and two 3$\times$3 convolutional layers, processing a concatenated input of current and historical discriminator outputs from the cache, as shown in Fig.~\ref{fig:discriminator}(a). The discriminator body includes a CQA block followed by strided convolution to handle detailed global features across domains. This integration of batch statistics, caching, and attention mechanisms significantly enhances the discriminator's ability to manage multi-resolution image data and stabilize the overall training (Figs.~\ref{fig:discriminator} b and c).
\subsection{Self-Supervised Pre-Training}
Self-supervised pre-training optimizes the initial weights of a network by engaging it in auxiliary tasks formulated from the inherent structure of the original data. In general, despite modifications in the cross-domain attention-guided discriminator, GAN training still suffers from mode collapse, where the generator repeatedly tends to produce a limited set of images. Thus, we opt to pre-train the generator of our Rawformer using self-supervised inpainting tasks to enhance initial weight optimization and mitigate this mode collapse problem.  \textit{ More details about the mode of pretraining is provided in the supplementary (Section 1) of the paper.}

%% file: sections_cr/experiments.tex
\section{Experiments}
We evaluated our method using two types of datasets : 1) raw-to-raw datasets, which provide unpaired raw images for training and aligned raw images from source and target cameras for assessing unsupervised raw-to-raw translation accuracy (Sec. \ref{sec:raw-to-raw-results}), and 2) raw-to-sRGB datasets, offering aligned raw and corresponding sRGB images for evaluating transformed raw image quality followed by learnable neural-based ISP rendering (Sec. \ref{sec:raw-to-srgb-results}). This evaluation mirrors real-world usage scenarios where raw images from a new camera are translated to the raw space of the camera used for ISP training. 
In all the experiments, raw images were demosaiced using the Menon algorithm \cite{menon2006demosaicing} and normalized after black-level subtraction. During our method's training, each raw image was segmented into non-overlapping crops of $224\times224$ pixels. Similarly, we used paired raw-sRGB non-overlapping crops of $224\times224$ pixels to train the camera-specific LAN ISP \cite{wirzberger2022lan} on each raw-to-sRGB dataset. 
\subsection{Datasets}
\label{sec:datasets}
We utilized two raw-to-raw datasets: 1) the Raw-to-Raw dataset~\cite{afifi2021semi} and 2) the NUS dataset~\cite{cheng2014illuminant}, and three raw-to-sRGB datasets: 1) the Zurich raw-to-RGB dataset~\cite{ignatov2019aim}, 2) the Samsung S7 dataset~\cite{schwartz2018deepisp}, and 3) the Mobile AIM21 dataset~\cite{Ignatov_2021_CVPR}. The Raw-to-Raw dataset~\cite{afifi2021semi} includes raw images from the Samsung Galaxy S9 and the iPhone X, featuring both unpaired and paired sets. The NUS dataset~\cite{cheng2014illuminant} comprises scenes captured by eight DSLR cameras. Following the evaluation in \cite{afifi2021semi}, we utilized the Nikon D5200 and Canon EOS 600D DSLR cameras from the NUS dataset \cite{cheng2014illuminant}. Aligned paired raw images were generated using the same approach as in the Raw-to-Raw dataset \cite{afifi2021semi}. In both datasets, paired raw images are dedicated for testing, while unpaired images are utilized for unsupervised training.

The Zurich raw-to-RGB dataset~\cite{ignatov2019aim} includes pairs of raw and sRGB images captured synchronously using a Huawei P20 smartphone camera and a Canon 5D Mark IV DSLR camera. The Samsung S7 dataset~\cite{schwartz2018deepisp} comprises paired raw and sRGB images captured by the Samsung S7 smartphone's main camera. The Mobile AIM21 dataset~\cite{Ignatov_2021_CVPR} includes images captured using a Sony IMX586 mobile sensor and a Fujifilm GFX100 DSLR, employing an advanced dense correspondence algorithm for matching due to imperfect alignment \cite{truong2021learning}.

\subsection{Training details}
\label{sec:training-details}
Inspired by the effective fine-tuning capabilities demonstrated by the pre-trained BERT model for the text completion task \cite{devlin2018bert}, we initially trained each generator network in a self-supervised fashion for image completion. Initially, the generator was trained on 32$\times$32 pixel patches extracted from training images, with 40\% of these patches being randomly masked out. The objective of the generator during this stage was to reconstruct the original unmasked image from its partially obscured version. Training was performed using the AdamW optimizer \cite{loshchilov2017decoupled} with betas set to (0.9, 0.99) and a weight decay of 0.05 for 500 epochs to minimize the following pixel-wise loss function:

\begin{equation}
    \mathcal{L}_{pixel-wise} = {L_1} + {(1-SSIM)} + {VGG}
    \label{eq:pixloss}
\end{equation}

\noindent where $L_1$, $SSIM$, $VGG$ refer to L1 loss, Structural Similarity Index (SSIM) \cite{wang2004image}, and VGG perceptual loss \cite{ledig2017photo} between original unmasked image and the reconstructed image. The learning rate was initialized at 0.005, and the learning schedule was managed using the cosine annealing warm restarts strategy \cite{loshchilov2016sgdr}. This self-supervised pre-training was shown to improve the raw mapping results (see Table \ref{tab:comparison2}).
 
Thereafter, the weights of both generators were fine-tuned for our unpaired raw-to-raw translation task for an additional 500 epochs, while simultaneously training the discriminator networks. We use the Adam optimizer~\cite{kingma2014adam} with betas set to (0.5, 0.99) and a learning rate of 0.0001 for the discriminator networks and 0.00005 for the generator networks. The discriminator networks were optimized to minimize the following loss functions:

\begin{equation}
    \mathcal{L}^{dis}_A = \ell_{gan}(D_A(G_{B \rightarrow A}(b)), 0) + \ell_{gan}(D_A(a), 1)
\end{equation}
\begin{equation}
    \mathcal{L}^{dis}_B =  \ell_{gan}(D_B(G_{A \rightarrow B}(a)), 0) +  \ell_{gan}(D_B(b), 1)
\end{equation}

\noindent where $a$ and $b$ represent images from camera $A$ and $B$, respectively, while $G_{B \rightarrow A}(b)$ and $G_{A \rightarrow B}(a)$ are the translated images produced by each generator network, respectively. The labels $0$ and $1$ represent ``fake'' and ``real'' raw images, respectively, and $\ell_{gan}(\cdot)$ computes the cross-entropy loss. 

During this stage, the weights of both generators were optimized to minimize the following loss function:

\begin{equation}
    \mathcal{L}^{gen} = \beta_1(\mathcal{L}^{gan}_A + \mathcal{L}^{gan}_B) + \beta_2(\mathcal{L}^{idt}_A + \mathcal{L}^{idt}_B) + \beta_3(\mathcal{L}^{cyc}_A + \mathcal{L}^{cyc}_B)
\end{equation}

\noindent where\hspace{3mm} $\beta_1=1$,\hspace{3mm} $\beta_2=10$,\hspace{3mm} $\beta_3=0.5$,\hspace{3mm}  $\mathcal{L}^{gan}_A =  \ell_{gan}(D_B(G_{A \rightarrow B}(a)), 1)$,\hspace{3mm}  $\mathcal{L}^{idt}_A = \mathcal{L}_{pixel-wise}(G_{B \rightarrow A}(a), a)$, and $\mathcal{L}^{cyc}_A =   \mathcal{L}_{pixel-wise}( G_{B \rightarrow A}(G_{A \rightarrow B}(a)), a)$. Note that at inference time, only a single generator network (e.g., $G_{A \rightarrow B}$) is required to translate raw images captured by the source camera (e.g., $A$) to the target camera (e.g., $B$).

\subsection{Raw-to-Raw Translation Results}
\label{sec:raw-to-raw-results}

\begin{figure*}[!t]
  \centering
  \includegraphics[width=0.9\linewidth]{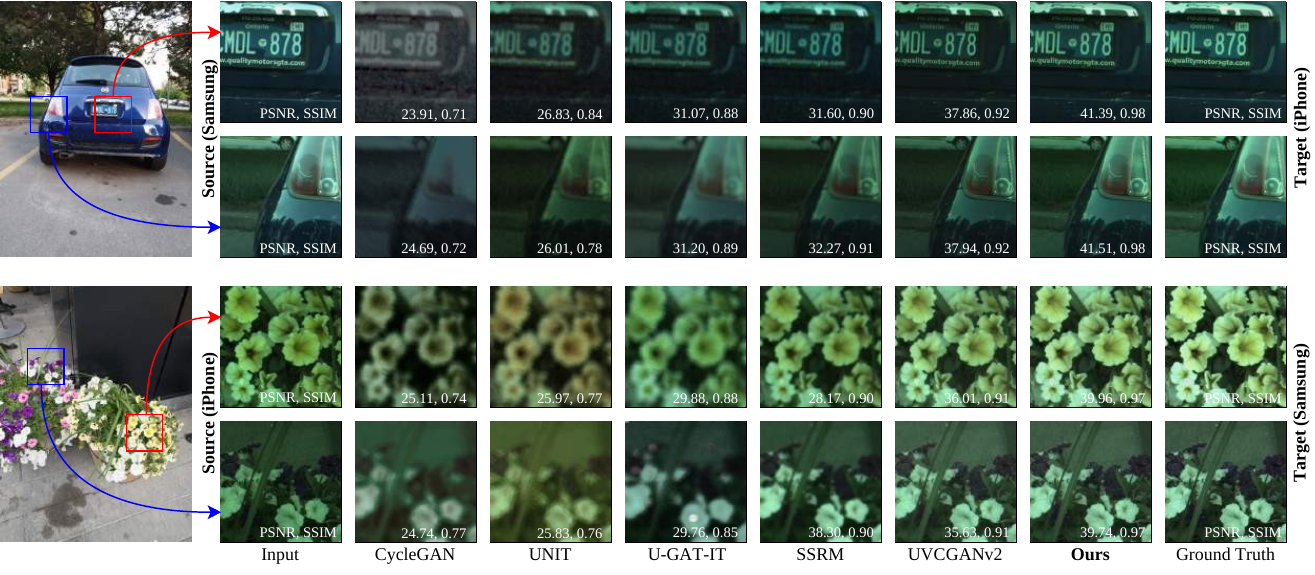}
  \caption{Shown are two images captured by Samsung S9 and iPhone X in sRGB (left) and two cropped patches from each image in raw (right). On the right, we show the input raw patch from the corresponding camera and the corresponding ground-truth raw patch from the other camera, along with the results by other methods. Our proposed Rawformer is better at preserving the domain consistent features.}
  \label{fig:vis-compare-results}
\end{figure*}

\begin{table}[!t]
\centering
\caption{\textbf{Raw-to-raw translation results on NUS dataset~\cite{cheng2014illuminant}.} The mapping results of Nikon D5200's raw images to Canon EOS 600D DSLR camera and vice versa using our method and other methods are shown. Here, $\ell_a = L_1$, $\ell_b = L_1 + VGG$, $\ell_c = \mathcal{L}_{pixel-wise}$ (Eq. \ref{eq:pixloss}), w, and w/o denote the results with and without training on a corresponding configuration, respectively. Additionally, we show the GFLOPs of each method for the 256×256 input image. \textit{Best results are in bold.}}
\label{tab:comparison2}

\resizebox{0.85\linewidth}{!}
{
\begin{tabular}{l|cccc|cccc|c}
\toprule
\multirow{2}{*}{Methods} & \multicolumn{4}{c}{Canon-to-Nikon} & \multicolumn{4}{c}{Nikon-to-Canon} \\
                        & PSNR$\uparrow$ & SSIM$\uparrow$ & MAE$\downarrow$ ($\times 1e-2$)    & $\Delta$E$\downarrow$    & PSNR$\uparrow$ & SSIM$\uparrow$& MAE$\downarrow$ ($\times 1e-2$)    & $\Delta$E$\downarrow$ & FLOPs (G)       \\
\midrule
CycleGAN~\cite{zhu2017unpaired}              & 27.32  & 0.83   & 3.21  & 15.40 & 26.81  & 0.82   & 3.33  & 11.49 & 36.8    \\
UNIT~\cite{liu2017unsupervised}               & 25.55  & 0.81   & 3.17  & 18.22 & 24.73  & 0.79   & 3.23  & 18.64  &  47.3  \\
U-GAT-IT~\cite{kim2019u}                & 29.41  & 0.87   & 3.02  & 7.12 & 27.86  & 0.85   & 3.29  & 8.01 &  62.9    \\
SSRM~\cite{afifi2021semi}                 & 32.36  & 0.93   & 2.41  & 6.21 & 30.81  & 0.93   & 2.47  & 5.95 &   37.1  \\
UVCGANv2~\cite{torbunov2023uvcgan2}              & 37.11  & 0.96   & 2.32  & 4.34 & 37.29  & 0.96   & 2.28  & 4.28 & 69.6     \\
\midrule
Ours (w/o self-supervised pre-training; $\ell_a$ loss)          & 37.13  & 0.96   & 2.29 & 4.65  & 37.27  & 0.96   & 2.31  & 4.46 & 52.7    \\
Ours (w/o self-supervised pre-training; $\ell_b$ loss)          & 37.94  & 0.97   & 2.21 & 3.96  & 37.99  & 0.96   & 2.23  & 3.93  &  52.7   \\
Ours (w/o self-supervised pre-training; $\ell_c$ loss)          & 38.73  & 0.97   & 2.19 & 3.64  & 38.26  & 0.97   & 2.22  & 3.72 &   52.7   \\
\textbf{Ours (w self-supervised pre-training; $\ell_c$ loss)}                    & \textbf{41.89} & \textbf{0.98} & \textbf{1.46}  & \textbf{2.04} & \textbf{41.37} & \textbf{0.98} & \textbf{1.41}  & \textbf{2.53} & 52.7 \\
\bottomrule
\end{tabular}
}
\end{table}

\begin{table}[!h]
\centering
\caption{\textbf{Raw-to-raw translation resuls on the Raw-to-Raw dataset~\cite{afifi2021semi}.} Shown are the mapping results of Samsung Galaxy S9's raw images to iPhone X and vice versa using our method and other methods. Additionally, the total number of parameters of each method is shown in millions (M), and inference time in milliseconds (ms) without optimization on NVIDIA GeForce RTX 4090 GPU. \textit{Results on inference time with optimization are in supplementary.}}
\label{tab:comparison}

\resizebox{0.85\linewidth}{!}
{
\begin{tabular}{l|cccc|cccc|c|c}
\toprule
\multirow{2}{*}{Methods} & \multicolumn{4}{c}{Samsung-to-iPhone} & \multicolumn{4}{c}{iPhone-to-Samsung} \\
                        & PSNR$\uparrow$ & SSIM$\uparrow$ & MAE$\downarrow$ ($\times 1e-2$)   & $\Delta$E$\downarrow$    & PSNR$\uparrow$ & SSIM$\uparrow$ & MAE$\downarrow$ ($\times 1e-2$)    & $\Delta$E$\downarrow$ & Params (M) & Time (ms)     \\
\midrule
CycleGAN~\cite{zhu2017unpaired}             & 24.63  & 0.71   & 5.40  & 14.71 & 24.37  & 0.73  & 4.34  & 12.88 & 21.7 & 19 \\
UNIT~\cite{liu2017unsupervised}              & 23.91  & 0.70   & 5.15  & 13.23 & 26.42  & 0.78  & 4.27  & 13.16 & 32.2 & 29 \\
U-GAT-IT~\cite{kim2019u}              & 28.22  & 0.89   & 3.11  & 5.98 & 31.04  & 0.89   & 2.39  & 5.04 & 34.4 & 35 \\
SSRM~\cite{afifi2021semi}              & 29.65  & 0.89   & 2.22  & 6.32 & 28.58  & 0.90   & 2.43  & 6.53 & \textbf{20.2} & \textbf{18} \\
UVCGANv2~\cite{torbunov2023uvcgan2}              & 36.32  & 0.94   & 2.09  & 4.21 & 36.46  & 0.92   & 2.02  & 4.73 & 32.6 & 31 \\
\midrule
\textbf{Ours}                    & \textbf{40.98} & \textbf{0.97} & \textbf{1.33}  & \textbf{2.09} & \textbf{41.48} & \textbf{0.98} & \textbf{1.21}  & \textbf{1.99} & {26.1} & 26  \\
\bottomrule
\end{tabular}
}
\end{table}

We compare our raw-to-raw translation results with those of the recent semi-supervised raw-to-raw mapping (SSRM) method \cite{afifi2021semi}. Additionally, we compare our method against several established unsupervised generic I2I methodologies, including UVCGANv2 \cite{torbunov2023uvcgan2}, U-GAT-IT \cite{kim2019u}, UNIT \cite{liu2017unsupervised}, and the classic CycleGAN \cite{zhu2017unpaired}. All other methods were trained on the same training data used for our method, for 500 epochs. Since SSRM is not entirely unsupervised, we incorporated 22 image pairs from the paired subset for its training, as discussed in \cite{afifi2021semi}. 
Fig. \ref{fig:vis-compare-results} shows qualitative comparisons on the Raw-to-Raw dataset \cite{afifi2021semi}, while Tables \ref{tab:comparison2} and \ref{tab:comparison} present the quantitative results on the NUS dataset \cite{cheng2014illuminant} and the Raw-to-Raw dataset, respectively. In Tables \ref{tab:comparison2} and \ref{tab:comparison}, we adopted the same quantitative metrics used in \cite{afifi2021semi}: peak signal-to-noise ratio (PSNR), SSIM, mean absolute error (MAE), and $\Delta$E 2000 \cite{sharma2005ciede2000}. As can be seen in both the tables, our Rawformer significantly outperforms the alternative methods across all the quantitative metrics by a large margin (around \textbf{$+$12} dB against SSRM \cite{afifi2021semi}). 
\subsubsection{Evaluation on Low-Light Images:}
Translating raw-to-raw for low-light images presents challenges due to various variations among cameras, including differences in raw normalization and noise characteristics \cite{afifi2021semi, abdelhamed2018high}. To further prove the efficacy of our method against other methods on low-light images, we selected a subset of 15 paired low-light images from the Raw-to-Raw dataset \cite{afifi2021semi}. The results on this subset are presented in Table \ref{tab:low}, where our proposed method significantly outperforms SSRM \cite{afifi2021semi}, and UVCGANv2 \cite{torbunov2023uvcgan2} for translating low-light raw images captured by an iPhone X to the Samsung S9's camera.

\begin{table}[!t]
\caption{\textbf{Raw-to-Raw dataset~\cite{afifi2021semi} results on low-light raw images (a subset).}}
\label{tab:low}
\parbox{.6\linewidth}{
\centering
  \includegraphics[width=0.9\linewidth]{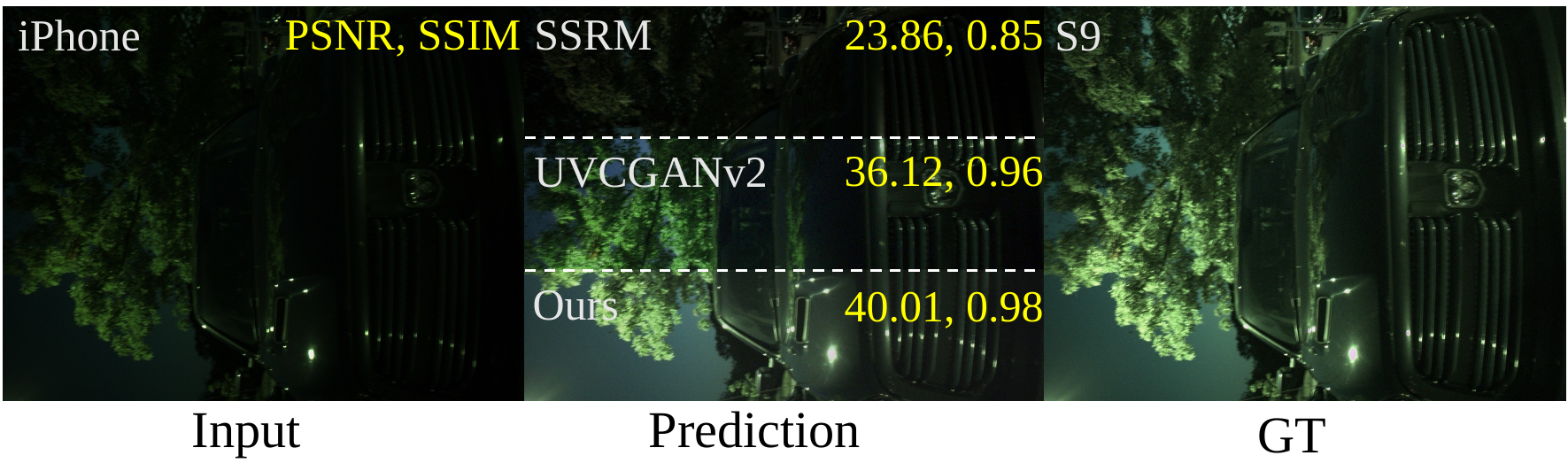}
}
\hfill
\parbox{.4\linewidth}{
\centering
\scalebox{0.7}{
\begin{tabular}{l|cc}
\toprule
Method & PSNR$\uparrow$ & SSIM$\uparrow$ \\
\midrule
SSRM  \cite{afifi2021semi}            & 23.87  & 0.85  \\
UVCGANv2  \cite{torbunov2023uvcgan2}        & 35.22  & 0.94  \\
\midrule
\textbf{Ours}           & \textbf{40.21} & \textbf{0.97} \\
\bottomrule
\end{tabular}
}}
\end{table}

\subsubsection{Ablation Studies:}
We conducted a set of ablation studies to validate the impact of the proposed components in the generator and discriminator networks. In this series of experiments, we trained our Rawformer on the NUS dataset \cite{cheng2014illuminant}, in addition to a baseline model trained using the same training settings as our Rawformer. The baseline model comprises a UNet architecture with channel and spatial attention \cite{torbunov2023uvcgan} for the generator, and a PatchGAN discriminator \cite{isola2017image}. As demonstrated in Table \ref{tab:study}, the integration of the proposed modules into the baseline consistently improves the network's capabilities, showcasing an enhancement in overall accuracy. We also conducted further ablation studies that explore various configurations of the discriminator, as shown in Table \ref{tab:study2}. The results clearly underscore the improvement by the inclusion of each proposed component in the discriminator network (attention, batch normalization, and discriminator head (Disc. head)) in improving the accuracy of our raw-to-raw translation.

\begin{table}[!t]
\centering
\caption{\textbf{Ablation results on the impact of various configurations of the generator model} on the NUS dataset \cite{cheng2014illuminant}. `CQA', `Style', `SPFN' and `CUp-CDown' refers to the condensed query attention, style modulator, scale perceptive feed-forward network, and the composite up and downsampler, respectively.}
\label{tab:study}
\resizebox{0.6\linewidth}{!}
{
\begin{tabular}{cccccc|cc|cc}
\toprule
\multicolumn{6}{c}{Modules} & \multicolumn{2}{c}{Canon-to-Nikon} & \multicolumn{2}{c}{Nikon-to-Canon} \\
Configuration & Baseline & CQA          & Style      & SPFN       & CUp-Cdown   & PSNR$\uparrow$& SSIM$\uparrow$ & PSNR$\uparrow$& SSIM$\uparrow$   \\
\midrule
$G_1$ & \checkmark &            &            &            &              & 37.23  & 0.96 & 37.31  & 0.96  \\
$G_2$ & \checkmark & \checkmark &            &            &              & 38.21  & 0.96 & 37.97  & 0.96  \\
$G_3$ & \checkmark & \checkmark & \checkmark &            &              & 39.08  & 0.97 & 38.99  & 0.96  \\
$G_4$ & \checkmark & \checkmark & \checkmark & \checkmark &              & 39.72  & 0.97 & 39.68  & 0.97  \\
$G_5$ & \checkmark & \checkmark & \checkmark & \checkmark & \checkmark   & \textbf{39.93}  & \textbf{0.97} & \textbf{39.86}  & \textbf{0.97}  \\
\bottomrule
\end{tabular}
}
\end{table}

\begin{table}[!t]
\centering
\caption{\textbf{Ablation results on the impact of various discriminator model configurations} on the NUS dataset~\cite{cheng2014illuminant}. `Baseline' refers to the $G_5$ config. from Tab.~\ref{tab:study}.}
\resizebox{0.65\linewidth}{!}
{
\label{tab:study2}
\begin{tabular}{ccccc|cc|cc}
\toprule
\multicolumn{5}{c}{Modules} & \multicolumn{2}{c}{Canon-to-Nikon} & \multicolumn{2}{c}{Nikon-to-Canon} \\
Configuration & Baseline & CQA & Batchnorm & Disc. head & PSNR$\uparrow$& SSIM$\uparrow$ & PSNR$\uparrow$& SSIM$\uparrow$   \\
\midrule
$D_1$ & \checkmark &            &            &            & 39.93  & 0.97 & 39.86  & 0.97  \\
$D_2$ & \checkmark & \checkmark &            &            & 40.12  & 0.97 & 40.09  & 0.97  \\
$D_3$ & \checkmark & \checkmark & \checkmark &            & 40.24  & 0.97 & 40.21  & 0.97  \\
$D_4$ & \checkmark & \checkmark & \checkmark & \checkmark & \textbf{41.89}  & \textbf{0.98} & \textbf{41.37}  & \textbf{0.98}  \\
\bottomrule
\end{tabular}
}
\end{table}

\subsection{Raw Translation for Learnable ISP}
\label{sec:raw-to-srgb-results}

\begin{figure}[!t]
  \centering
  \includegraphics[width=0.85\linewidth]{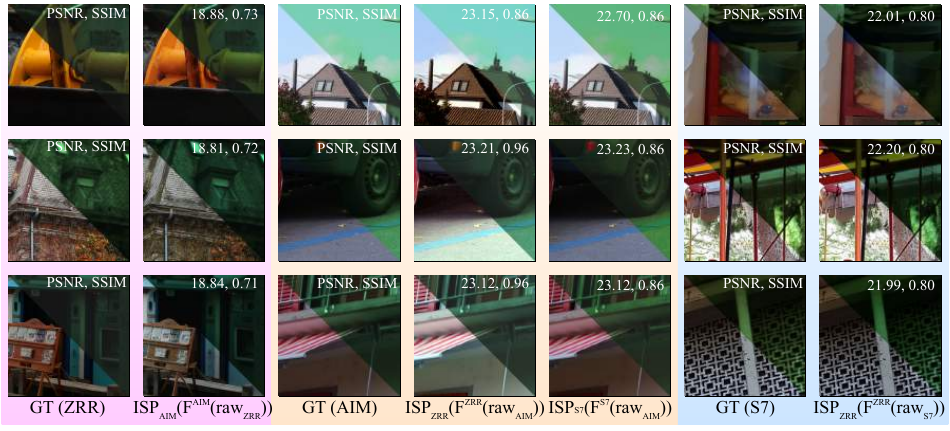}
  \caption{\textbf{ISP rendering results with our raw translation on various datasets.} Each set includes ground-truth (GT) raw-sRGB paired images and LAN ISP~\cite{wirzberger2022lan} results on mapped raw images from different cameras.  
  S7, ZRR, and AIM stand for the Samsung S7 ISP dataset~\cite{schwartz2018deepisp}, Zurich raw-to-RGB dataset~\cite{ignatov2019aim}, and Mobile AIM21 dataset~\cite{Ignatov_2021_CVPR}, respectively. The shown results are consistent with the ground-truth, demonstrating the proficiency of our model.} 
  \label{fig:vis-isp}
\end{figure}

To comprehend the practical application of our model in real-world scenarios, we utilized three datasets: Zurich raw-to-RGB, Samsung S7 ISP, and Mobile AIM21, encompassing various source and target domain combinations. Specifically, for each dataset, we trained the LAN ISP \cite{wirzberger2022lan} on a single paired raw-to-sRGB dataset, where raw images originated from a single camera. Then, we transferred all raw images from other cameras in the other datasets using our Rawformer before rendering with the trained LAN model. We report the PSNR and SSIM of the rendered sRGB images compared to the ground-truth sRGB images for each set as shown in Table \ref{tab:da}. This includes the results of \textit{camera-specific} LAN ISP models trained and tested on the same datasets. The camera-specific results represent the optimal results achievable by the neural-based ISP, which requires a paired raw-SRGB dataset for training on each camera model. As can be seen, our unsupervised raw-to-raw translation method demonstrates robust improvement, exhibiting only a marginal 3\% reduction in accuracy compared to the camera-specific LAN model trained and tested on the same camera. This underscores the effectiveness of our method in addressing real-world cross-domain broadcasting challenges. See Fig. \ref{fig:vis-isp} for qualitative examples.
\begin{table}[!t]
\caption{\textbf{Cross-camera raw-to-sRGB results (PSNR/SSIM)} comparing: a) UVCGANv2 \cite{torbunov2023uvcgan2} with b) our Rawformer. Each entry compares ground-truth sRGB images generated by the function: $ISP_B(F^{B}(raw_A))$, where $F^{B}$ denotes translation of raw images, $raw_A$, captured by camera $A$ to camera $B$. $ISP_B(\cdot)$ produces sRGB images using the LAN ISP \cite{wirzberger2022lan} trained on raw images from camera $B$. Diagonal results (in \textcolor{red}{red}) represent the \textit{camera-specific} ISP case and are the sRGB images produced by the ISP model tested on the same camera used for training.  Best cross-camera results are in bold.}
\label{tab:da}
\centering
\resizebox{0.7\linewidth}{!}
{

\begin{tabular}{lccc|ccc}
\toprule
\multicolumn{1}{c}{} & \multicolumn{3}{c|}{a) UVCGANv2 \cite{torbunov2023uvcgan2}$+$ LAN ISP \cite{wirzberger2022lan}} & \multicolumn{3}{c}{b) Our Rawformer$+$ LAN ISP \cite{wirzberger2022lan}} \\ \cline{2-7} 
\multicolumn{1}{c}{} & \multicolumn{3}{c|}{Testing camera B} & \multicolumn{3}{c}{Testing camera B} \\ \cline{2-7} 
\multicolumn{1}{l|}{Training camera A} & ZRR \cite{ignatov2019aim} & AIM21~\cite{Ignatov_2021_CVPR} & S7  \cite{schwartz2018deepisp} & ZRR \cite{ignatov2019aim} & AIM21~\cite{Ignatov_2021_CVPR} & S7  \cite{schwartz2018deepisp} \\ \hline
\multicolumn{1}{l|}{ZRR \cite{ignatov2019aim}} &  \textcolor{red}{19.46/0.73} & 17.72/0.70  & 17.81/0.70  & \textcolor{red}{19.46/0.73} & \textbf{18.97}/\textbf{0.72}  & \textbf{19.02}/\textbf{0.72}   \\
\multicolumn{1}{l|}{AIM21~\cite{Ignatov_2021_CVPR}} & 22.14/0.79 & \textcolor{red}{23.48/0.87} & 22.39/0.79 & \textbf{23.25}/\textbf{0.86} & \textcolor{red}{23.48/0.87} & \textbf{23.08}/\textbf{0.86}  \\
\multicolumn{1}{l|}{S7 \cite{schwartz2018deepisp}} &  21.07/\textbf{0.80} & 20.93/0.79  & \textcolor{red}{22.16/0.81} & \textbf{22.08}/\textbf{0.80} & \textbf{22.03}/\textbf{0.80}  & \textcolor{red}{22.16/0.81}\\
\bottomrule
\end{tabular}
}
\end{table}

\subsection{Comparison with Few-Shot Domain Adaptation}

We further compared our Rawformer against Prabhakar's Few-Shot Domain Adaptation (PDA) \cite{prabhakar2023few}. Specifically, we conducted a 10-shot domain adaptation from the Samsung S7 ISP dataset's camera \cite{schwartz2018deepisp} to the Zurich raw-to-RGB dataset's camera \cite{ignatov2019aim}, and vice-versa. The PDA model underwent initial training for 10 epochs using the Samsung S7 ISP dataset. Subsequently, domain adaptation was performed on 10 images from the Zurich raw-to-RGB dataset's camera over 4 epochs. This process resulted in performance metrics of 21.07 dB, and 17.81 dB in terms of PSNR and 0.80 and 0.70 for SSIM for our Rawformer, representing approximately 11\% superior performance compared to PDA (see Table~\ref{tab:p-da}).

\begin{table}[!h]
  \centering
  \caption{\textbf{Comparison with PDA~\cite{prabhakar2023few}.} S7 and ZRR denote the Samsung S7 ISP~\cite{schwartz2018deepisp} and the Zurich Raw-to-RGB~\cite{ignatov2019aim} datasets, respectively. Results are produced after translating each raw image to the target camera using PDA and our Rawformer, subsequently rendering with a LAN ISP \cite{wirzberger2022lan} trained on the target camera's raw images.}
  \label{tab:p-da}
  {
  \scalebox{0.6}{
\begin{tabular}{l|cccc|cccc}
\toprule
\multirow{2}{*}{Method} & \multicolumn{4}{c|}{S7-to-ZRR} & \multicolumn{4}{c}{ZRR-to-S7} \\
                        & PSNR$\uparrow$ & SSIM$\uparrow$ & MAE$\downarrow$ ($\times 1e-2$)   & $\Delta$E$\downarrow$    & PSNR$\uparrow$ & SSIM$\uparrow$ & MAE$\downarrow$ ($\times 1e-2$)    & $\Delta$E$\downarrow$       \\
\midrule
PDA~\cite{prabhakar2023few}            & 20.33  & 0.79   & {9.41}  & 13.99 & 17.12  & 0.69   & 18.03  & 15.86    \\
\midrule
\textbf{Ours}           & \textbf{22.08} & \textbf{0.80} & \textbf{8.18}  & \textbf{10.27} & \textbf{19.02} & \textbf{0.72} & \textbf{15.27}  & \textbf{11.74}      \\
\bottomrule
\end{tabular}}}

\end{table}

%% file: sections_cr/conclusion.tex
\section{Conclusion and Future Work}
In this work, we presented a new approach, Rawformer, for raw-to-raw image translation using unpaired data.  Rawformer sets a new standard for performance on real camera datasets, showcasing its superior capabilities in raw-to-raw translation when measured against other
 alternatives. The proposed method reduces the need for the extensive collection of raw-to-sRGB paired datasets to re-train learnable ISPs for new cameras, thereby decreasing associated costs. Our results indicate that existing raw-to-sRGB paired datasets can reliably train neural-based learnable ISPs, which can then be utilized to render raw images taken by unseen cameras, lacking such unpaired data. We believe that our approach provides valuable insights for learnable ISP, raw image translation research, and applications. Our method (without optimization) runs at 26 milliseconds per frame on GPU ($\sim$1 second on CPU), which may be impractical for real-time rendering in camera previews on devices with limited computational power. Future work will focus on introducing lighter models capable of real-time performance on CPU.

%% file: sections_cr/supp_materials.tex
In this supplementary material, we first discuss the holistic architecture of our proposed Rawformer along with the experimental details in Sec. \ref{arch}. Then, we delve into the details of the style modulator along with the demonstration of its efficacy in the restoration of the overall structural fidelity of the translated images in Sec. \ref{style}. Next, we present some additional ablation studies and inference time comparison to prove the effectiveness of the proposed components in Sec. \ref{ablation}. Lastly, we provide additional results, including mappings between DSLR and mobile phone cameras, as well as additional visual results, in Sec. \ref{visual}. 
\section{Architecture and Experimental Details}
\label{arch}
\begin{figure*}[h]
  \centering
  \includegraphics[width=1\linewidth]{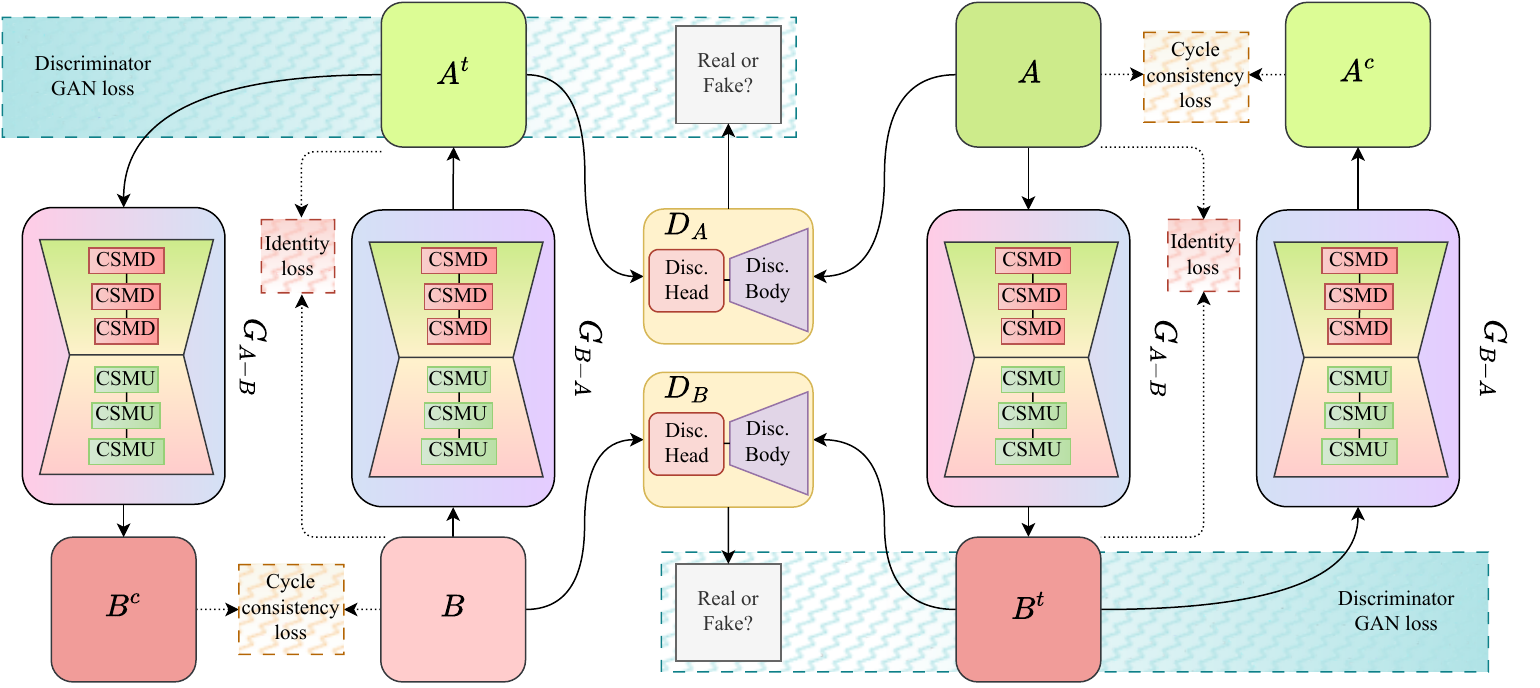}
  \caption{Overview of the proposed architecture and training flow. $A^t$ and $B^t$ refer to translated images used by the discriminator loss, while $A^c$ and $B^c$ refer to the produced images used by the cycle consistency loss.}
  \label{supp-fig:overview}
\end{figure*}

In the main paper, we introduced Rawformer, a fully unsupervised framework for raw-to-raw translation. The overview of the proposed framework and the training flow are illustrated in Fig. \ref{supp-fig:overview}. As can be seen, we train two generator networks, $G_{A-B}$ and $G_{B-A}$, alongside two discriminator networks, $D_A$ and $D_B$, utilizing cycle consistency loss, identity loss, and discriminator loss. Our framework accurately maps images from domain $A$ to domain $B$ through a fully unsupervised training scheme. 

\subsection{Pre-training of Generator Networks}
The pre-training phase of the generator network, essential for initializing the weights effectively, involves a self-supervised image inpainting task spanning 500 epochs. This task is designed to enhance the network's detail preservation capabilities in input images. Specifically, the network is trained on 32$\times$32 pixel patches from the source dataset comprising of images of size 256$\times$256, with 40\% of these patches randomly masked. The masking is conducted by zeroing out the pixel values. The objective of the generator is to reconstruct the original image from its partially obscured version by minimizing a pixel-wise loss function, as discussed in the main paper. The pre-training is performed with batch size of 16 and we apply random horizontal and vertical flips on both the considered datasets. This pretraining phase employs an AdamW optimizer with betas (0.9, 0.99), a weight decay of 0.05, an initial learning rate of 0.005, and a learning schedule managed through the Cosine Annealing Warm Restarts strategy. 

\subsection{Raw Translation Training}

After pre-training, the model enters the Generative Adversarial Network (GAN) training phase, which lasts for 500 epochs and focuses on unpaired image translation. This phase employs a pixel-wise loss function for both the generator and discriminator components, with the discriminator optimized using the Adam optimizer with betas (0.5, 0.99) and a learning rate of 0.0001. The generator uses the same optimizer but with a slightly lower learning rate of 0.00005. The batch size is kept to 1 and data augmentations, including  random horizontal flips and vertical flips, are further applied on the dataset in this phase as well. The overall size of image caches is set to 3 and the overall batch head has 4 samples to compute the batch statistics. The overall training process is illustrated in Fig. \ref{supp-fig:overview}.


\section{Style Modulator}
\label{style}

\begin{figure}[t]
\centering
\includegraphics[width=0.6\linewidth]{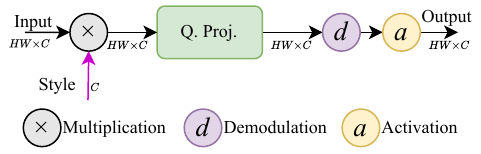}
\caption{\textbf{Details of the proposed style modulation process.}}
\label{supp-fig:style-arch}
\end{figure}

To augment the capability of the generator, we have expanded its functionality to deduce the correct target style for every input image, using a Vision Transformer (ViT) \cite{torbunov2023uvcgan2}. Thereafter, we modulate the decoding part of the generator with the obtained target style for markedly boosting its expressive power \cite{karras2020analyzing}. The main technique of style modulation is depicted in Figure \ref{supp-fig:style-arch}. 

In particular, at the generator's bottleneck, the image is encoded into a series of tokens for the ViT network. We enhance this series with an extra trainable style token $S$, which, after processing through the ViT, encapsulates the latent style of the image. For each layer in the Rawformer's decoding section, we derive a unique style vector $s_i$ from $S$ through trainable linear transformations.

\begin{figure}[t]
\centering
\includegraphics[width=0.9\linewidth]{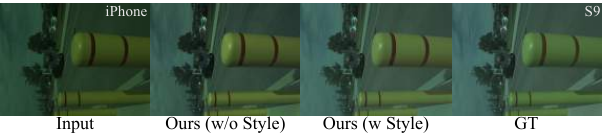}
\caption{\textbf{Enhancement in translation accuracy by the style modulator.} The improvements are demonstrated on the Raw-to-Raw dataset~\cite{afifi2021semi} (from iPhone X to Samsung S9), showcasing the positive impact of the style modulator on translation accuracy.}
\label{supp-fig:style}
\end{figure}

The process of style modulation \cite{karras2020analyzing} effectively adjusts the weights $w_{i,j,h,w}$ of the $Q$ vector with the designated style vector $s_i$, resulting in modulated weights:
\begin{equation}
w'_{i,j,h,w} = s_i \times w_{i,j,h,w},
\end{equation}
where $i, j$ denote the input and output feature maps, respectively, and $h, w$ represent spatial dimensions. To maintain the activation magnitudes, the modulated weights, $w'_{i,j,h,w}$, are subjected to demodulation, renormalizing the convolutional $Q$ vector weights as:
\begin{equation}
w''_{i,j,h,w} = \frac{w'_{i,j,h,w}}{\sqrt{\sum_{h,w} (w'_{i,j,h,w})^2 + \epsilon}},
\end{equation}
with $\epsilon$ being a minimal value to avoid numerical issues.
Figure \ref{supp-fig:style} illustrates that the introduced style modulator significantly enhances translation precision.

\section{Inference time and Additional Ablation studies}
\label{ablation}

\subsection{Inference time}

\begin{table}[t]
\centering
\caption{\textbf{Quantitative results of optimized models.} Shown results are for Samsung-to-iPhone mapping (using the Raw-to-Raw dataset \cite{afifi2021semi}) on different hardware platforms, along with the inference time in milliseconds (ms). The results show that our proposed Rawformer holds promise for integration into the mobile devices.}
\label{supp-tab:inference}
\begin{tabular}{l|cc|c|cccc}
\toprule
Device & Dtype & Framework & Time (ms) & PSNR & SSIM & MAE & $\Delta$E     \\
\midrule
CPU Intel i9-12900K & fp32  & PyTorch & 526 & \textbf{40.98} & \textbf{0.97} & \textbf{0.01}  & \textbf{2.09} \\
CPU Intel i9-12900K & int8  & OpenVINO & 179 & 37.21 & 0.95 & 0.03  & 5.14 \\
GPU NVIDIA RTX 4090 & fp32  & PyTorch & 26 & \textbf{40.98} & \textbf{0.97} & \textbf{0.01}  & \textbf{2.09}\\
GPU NVIDIA RTX 4090 & fp16  & TensorRT & \textbf{18} & 40.92 & \textbf{0.97} & \textbf{0.01}  & 2.11 \\
Google Coral Edge TPU & int8  & TF Lite & \underline{68}  & 37.20 & 0.95 & 0.03  & 5.21 \\
\bottomrule
\end{tabular}
\end{table}

In the main paper, we presented the inference time of our model without optimization. Here, we present the results of various optimized versions of our model for raw translation. Specifically, we applied model post-training quantization to our trained Rawformer, which was trained to map Samsung S9's raw images to iPhone X's camera raw space. We report the results of float16, float32 and int8 quantization on different hardware platforms in Table \ref{supp-tab:inference}. It is worth noting that we achieve nearly identical accuracy with float16 conversion, running at approximately 18 milliseconds on an NVIDIA RTX 4090 GPU.

\subsection{Ablation Study}
In the main paper, we presented several ablation studies conducted to validate the decisions made in the proposed design of Rawformer. Here, we present additional ablation experiments performed to further validate the operations of the major components in our Rawformer: condensed query attention (CQA), contextual-scale aware upsampler (CSAU) block and the contextual-scale aware downsampler (CSAD) blocks. Tables \ref{supp-tab:study_kqv}, and \ref{supp-tab:study_pool} demonstrates the impact of incorporating the spatial compression (where and how), on the attention vectors of CQA block. The results of the ablation studies shown in  Tables \ref{supp-tab:study_up}, and \ref{supp-tab:study_updown} proves the efficacy of our designed composite upsamplers and downsamplers. Additionally, to prove that the different hierarchical levels in our proposed SPFN module offer a more intuitive approach for the impact of capturing local image structure across different scales, we provided an ablation study as shown in Table \ref{ans-tab:spfn}. All these ablation studies clearly reveal, that the proposed design with the inclusion of all the components achieves the best results across all quantitative metrics.

\begin{table}[t]
\centering
\caption{\textbf{Ablation results of the impact of the spatial conpression operation on different vectors of attention} on the NUS dataset~\cite{cheng2014illuminant}.  `Q', `K', and `V' indicate the use of the $Q$, $K$, and $V$ vectors in the CQA block. `Ours' represents the proposed design discussed in the main paper. The shown results exhibits the proficiency of applying the spatial compression/condensation operation on the query vectors.}

\label{supp-tab:study_kqv}
\begin{tabular}{c|cc|cc}
\toprule
              & \multicolumn{2}{c}{Canon-to-Nikon} & \multicolumn{2}{c}{Nikon-to-Canon} \\
Spatial Compression operation & PSNR$\uparrow$& SSIM$\uparrow$ & PSNR$\uparrow$& SSIM$\uparrow$   \\
\midrule
K          & 41.12  & 0.98 & 41.09  & 0.98  \\
V          & 41.14  & 0.98 & 41.11  & 0.98  \\
\textbf{Q (Ours)}   & \textbf{41.89}  & \textbf{0.98} & \textbf{41.37}  & \textbf{0.98}  \\
\bottomrule
\end{tabular}
\end{table}

\begin{table}[t]
\centering
\caption{\textbf{Ablation results on the impact of different query projection operations for the condensed query attention block}. Here `Patch merging' involves splitting the incoming image feature into patches and then merging across the channel dimension, and  `Ours' represents the proposed design discussed in the main paper. The shown results clearly reveal that using average pooling and linear projection, helps in refining the overall results.}

\label{supp-tab:study_pool}
\begin{tabular}{c|cc|cc}
\toprule
              & \multicolumn{2}{c}{Canon-to-Nikon} & \multicolumn{2}{c}{Nikon-to-Canon} \\
Query Projection Operation & PSNR$\uparrow$& SSIM$\uparrow$ & PSNR$\uparrow$& SSIM$\uparrow$   \\
\midrule

Depthwise Conv (stride=2) & 40.13  & 0.97 & 40.12  & 0.97  \\
Conv  (stride=2)         & 39.52  & 0.97 & 39.48  & 0.97  \\
Patch merging  & 38.83  & 0.96 & 38.73  & 0.96  \\
MaxPool + Linear Projection     & 41.09  & 0.98 & 41.03  & 0.98  \\
\textbf{AvgPool + Linear projection} & \textbf{41.89}  & \textbf{0.98} & \textbf{41.37}  & \textbf{0.98}  \\
\bottomrule
\end{tabular}
\end{table}

\begin{table}[t]
\centering
\caption{\textbf{Ablation results on the impact of different generator model configurations} on the NUS dataset~\cite{cheng2014illuminant}. `CUp' indicates the upsampling block in the contextual-scale aware upsampler (CSAU) block, while `Deconv' refers to the classical deconvolution block. `Ours' represents the proposed design discussed in the main paper. The shown results demonstrate the merits of deploying hybrid upsampling in the overall design.}

\label{supp-tab:study_up}
\begin{tabular}{c|cc|cc}
\toprule
              & \multicolumn{2}{c}{Canon-to-Nikon} & \multicolumn{2}{c}{Nikon-to-Canon} \\
Configuration & PSNR$\uparrow$& SSIM$\uparrow$ & PSNR$\uparrow$& SSIM$\uparrow$   \\
\midrule
Deconv     & 40.93  & 0.97 & 40.84  & 0.97  \\
CUp (Ours)  & \textbf{41.89}  & \textbf{0.98} & \textbf{41.37}  & \textbf{0.98}  \\
\bottomrule
\end{tabular}
\end{table}

\begin{table}[t]
\centering
\caption{\textbf{Ablation results on the impact of different generator model configurations} on the NUS dataset~\cite{cheng2014illuminant}. `CDown' indicates downsampling in the contextual-scale aware downsampler (CSAD) block, while `Conv' signifies the simple convolution block with stride 2 for downsampling. `Ours' represents the proposed design discussed in the main paper.}

\label{supp-tab:study_updown}
\begin{tabular}{c|cc|cc}
\toprule
              & \multicolumn{2}{c}{Canon-to-Nikon} & \multicolumn{2}{c}{Nikon-to-Canon} \\
Configuration & PSNR$\uparrow$& SSIM$\uparrow$ & PSNR$\uparrow$& SSIM$\uparrow$   \\
\midrule
Conv (stride 2) & 40.14  & 0.97 & 40.07  & 0.97  \\
CDown (Ours) & \textbf{41.89}  & \textbf{0.98} & \textbf{41.37}  & \textbf{0.98}  \\
\bottomrule
\end{tabular}
\end{table}

\begin{table}[h]

\centering
\caption{\textbf{Ablation results of different modifications of SPFN.}}
\label{ans-tab:spfn}
\begin{tabular}{l|cccc}
\toprule
Task & PSNR & SSIM & MAE ($\times 1e-2$) & $\Delta$E     \\
\midrule
Multi-scale SPFN & 41.89 & 0.98 & 1.26  &  2.04 \\
Single-scale SPFN & 41.11 & 0.97 & 1.70  & 2.57 \\
\bottomrule
\end{tabular}
\end{table}

\section{Comparison with SoTA approaches}
Our method represents the first practical and precise solution to this challenge, relying solely on unpaired sets of raw images. While our network design draws inspiration from existing techniques like channel dependencies from SENet \cite{hu2018squeeze} and depthwise convolutions from MobileNet \cite{howard2017mobilenets}, our goal is to develop an efficient network capable of achieving precise unpaired image translation with minimal resource requirements. Given the typical deployment of such tasks on resource-constrained devices, our modifications aim to not only reduce resource demands compared to alternatives mentioned above, but also to enhance accuracy. As demonstrated in the main paper (Tables 1-3), our design outperforms several alternative methods (e.g., UVCGANv2). Even when substituting certain blocks in our design with alternative ones, our proposed design achieves better results with reduced computations. For instance, our discriminator utilizes an attention mechanism with caching, yielding improved results compared to the UVCGANv2 discriminator: 41.89 dB~vs.~40.51 dB on the Canon-to-Nikon set (CNS). Our efficient CQA block, which performs spatial compression/condensation operations on query vectors, achieves better results compared to traditional self-attention: 41.89 dB~vs.~39.93 dB on CNS with a 50.7\% reduction in FLOPs (also see Table~2 in supp. materials). Using our single resolution SPFN block leads to improved results compared to mutli-resolution HRNet \cite{yu2021lite} feedforward block : 41.89 dB~vs.~41.77 dB on CNS with $\sim$87\% reduction in FLOPs. We will further refine our main contributions and clarify our design motivation, as suggested by R1\& R4, respectively thus eventually highlighting the ability of accurate raw-to-raw with unpaired data.

\section{Additional Results}
\label{visual}
\begin{figure*}[!t]
  \centering
  \includegraphics[width=1\linewidth]{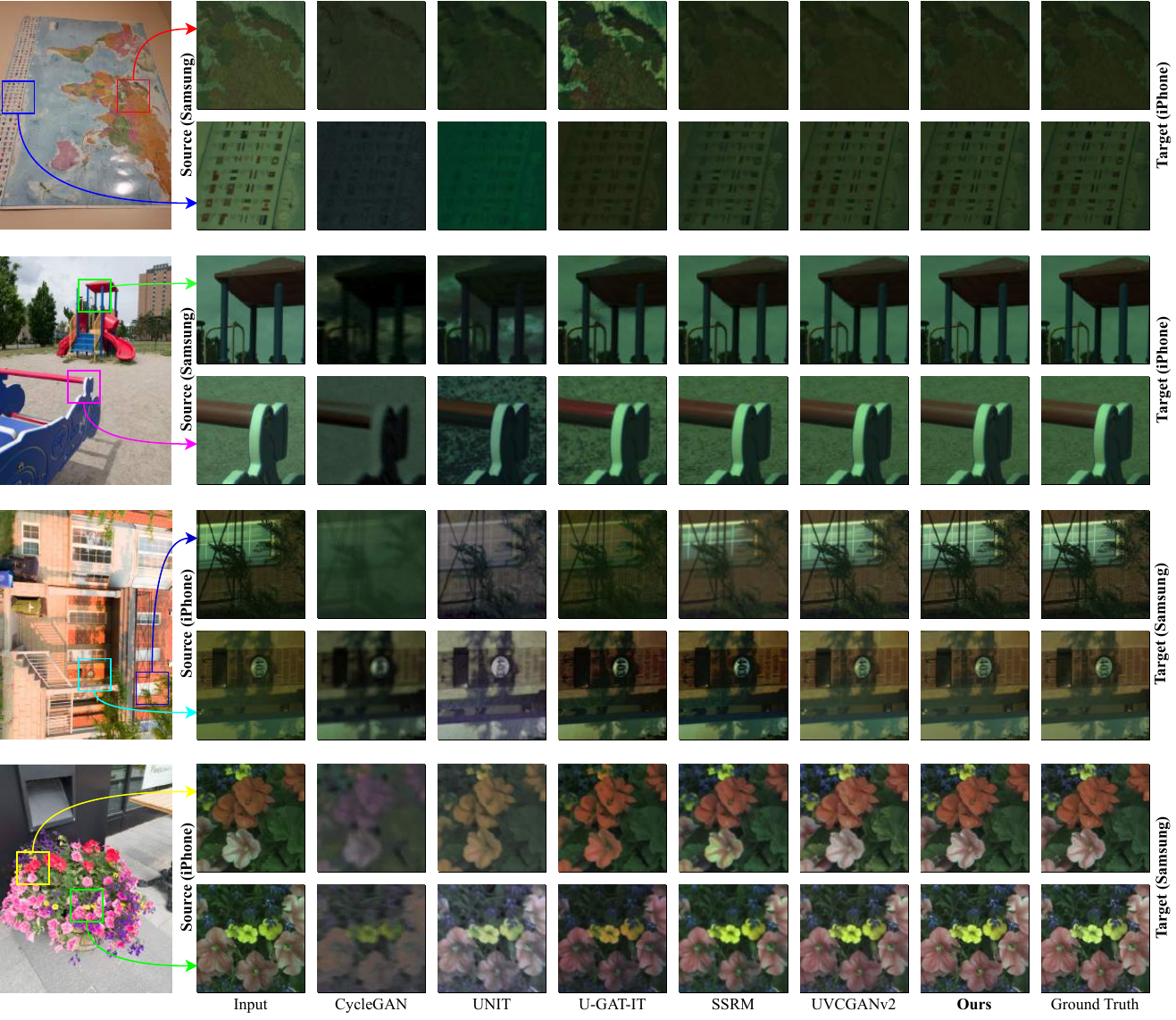}
  \caption{\textbf{Qualitative results of raw translation on the Raw-to-Raw dataset \cite{afifi2021semi}.} Shown are images captured by Samsung S9 and iPhone X in sRGB (left) and two cropped patches from each image in raw (right). On the right, we show the input raw patch from the corresponding camera and the corresponding ground-truth raw patch from the other camera, along with the results by other methods. Our proposed Rawformer is better at preserving the domain consistent features.}
  \label{supp-fig:vis-compare-results}
\end{figure*}

The experiments presented in the main paper focus on the raw translation of raw images captured by various mobile phone cameras, representing a real-world scenario and demonstrating one of the most promising applications of the proposed method---reducing costs associated with the development of mobile phone camera's neural-based ISP for new camera models. Here, we conducted experiments where we examined the raw mapping between DSLR and mobile phone cameras. Specifically, we trained our method, along with other unsupervised methods \cite{zhu2017unpaired, liu2017unsupervised}, to map between the Canon EOS 600D DSLR camera (from the NUS dataset \cite{cheng2014illuminant}) and the main camera of the Huawei P20 smartphone (from the Zurich raw-to-RGB dataset \cite{ignatov2019aim}). Additionally, we trained a LAN neural-based ISP \cite{wirzberger2022lan} on the target camera. In Table \ref{supp-tab:dslr-to-raw-comparison}, we present quantitative results obtained by rendering raw images mapped using our method and other methods, utilizing the pre-trained ISP. As demonstrated, our approach yields superior raw translation results compared to alternative methods, as evidenced by the quality of the rendered sRGB images in comparison to the ground-truth sRGB images.

\begin{figure}[t]
\centering
\includegraphics[width=0.7\linewidth]{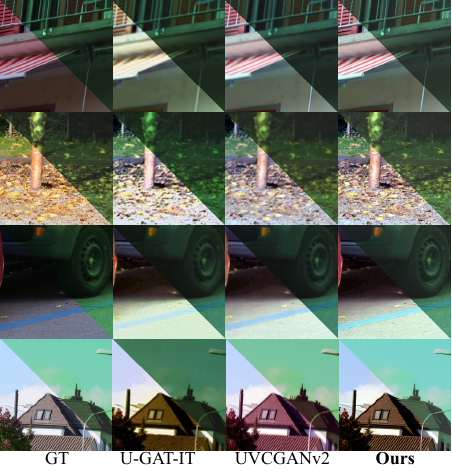}
\caption{\textbf{Qualitative comparisons on raw translation and ISP rendering.} We show the ground-truth (GT) raw/sRGB images from the the Mobile AIM21 dataset (Sony IMX586) \cite{Ignatov_2021_CVPR}, alongside the corresponding mapped raw images to the Zurich raw-to-RGB dataset (Huawei P20 smartphone's main camera) \cite{ignatov2019aim} generated using various methods, including ours. Additionally, we show the rendered sRGB images by processing each mapped raw image using a neural-based ISP~\cite{wirzberger2022lan} trained to render raw images from the Zurich dataset source camera (i.e., the Huawei P20 smartphone camera).}
\label{supp-fig:vis-compare-results2} 
\end{figure}

\begin{figure}[!t]
  \centering
  \includegraphics[width=\linewidth]{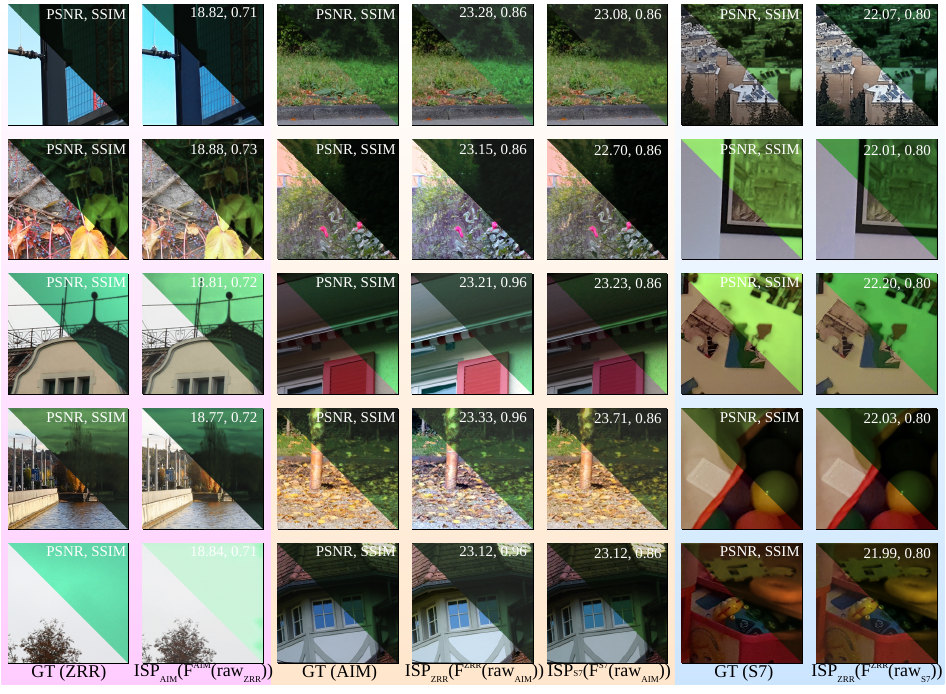}
  \caption{\textbf{ISP rendering results with our raw translation on various datasets.} Each set includes ground-truth (GT) raw-sRGB paired images and LAN ISP~\cite{wirzberger2022lan} results on mapped raw images from different cameras. $F^{y}$ represents our Rawformer trained to map raw images, $\texttt{raw}_x$, from a source camera, $x$, to target camera, $y$. $ISP{y}$ denotes LAN ISP~\cite{wirzberger2022lan} trained on raw images from camera $y$. 
  S7, ZRR, and AIM stand for the Samsung S7 ISP dataset~\cite{schwartz2018deepisp}, Zurich raw-to-RGB dataset~\cite{ignatov2019aim}, and Mobile AIM21 dataset~\cite{Ignatov_2021_CVPR}, respectively. The shown results are consistent with the ground-truth, demonstrating the proficiency of our model.}  
  \label{supp-fig:vis-isp}
\end{figure}

\begin{table}[!t]
\centering
\caption{\textbf{Translation results for mapping between raw images from DSLR and mobile phone cameras}. The results are on the NUS dataset~\cite{cheng2014illuminant} and the Zurich raw-to-RGB dataset (ZRR) \cite{ignatov2019aim}. Specifically, the mapping results of  ZRR raw images (captured by the Huawei P20 smartphone camera) to the Canon EOS 600D DSLR camera, and vice versa, are shown. Both our method and other techniques are compared. \textit{Best results are highlighted in bold.}}
\label{supp-tab:dslr-to-raw-comparison}
\begin{tabular}{l|cc|cc}
\toprule
\multirow{2}{*}{Methods} & \multicolumn{2}{c}{Canon-to-ZRR} & \multicolumn{2}{c}{ZRR-to-Canon} \\
                        & PSNR$\uparrow$ & SSIM$\uparrow$  & PSNR$\uparrow$ & SSIM$\uparrow$\\
\midrule
CycleGAN~\cite{zhu2017unpaired}              & 12.63  & 0.54  & 12.81  & 0.58   \\
UNIT~\cite{liu2017unsupervised}              & 14.91  & 0.67  & 17.73  & 0.70   \\
UVCGANv2~\cite{torbunov2023uvcgan2}          & 17.32  & 0.71  & 22.29  & 0.87   \\
\textbf{Ours}                    & \textbf{18.71} & \textbf{0.72} & \textbf{24.35} & \textbf{0.89} \\
\bottomrule
\end{tabular}
\end{table}

Additional qualitative raw translation results are shown in Figs. \ref{supp-fig:vis-compare-results} and \ref{supp-fig:vis-compare-results2}. In Fig. \ref{supp-fig:vis-compare-results2}, we also show the results of mapped raw images (transformed from the Sony IMX586 camera to the main camera of the Huawei P20 smartphone) after rendering using a pre-trained LAN neural-based ISP \cite{wirzberger2022lan}, which was trained on the raw space of the target camera. Specifically, the ISP was trained to render images captured by the Huawei P20 smartphone's main camera. As can be seen in Figs. \ref{supp-fig:vis-compare-results} and \ref{supp-fig:vis-compare-results2}, our method achieves superior raw translation, resulting in visually enhanced sRGB images when rendered using the pre-trained ISP compared to other alternative methods. Lastly, Fig. \ref{supp-fig:vis-isp} shows additional qualitative results of our raw mapping and the rendered sRGB images using trained ISP \cite{wirzberger2022lan} on the target camera raw space.


Additionally, we conducted experiments to test the impact of our proposed method on rendering images using Adobe Lightroom. For this experiment, we used the Raw-to-Raw dataset \cite{afifi2021semi}. We compared raw images captured by a Samsung smartphone, rendered by Adobe Lightroom with metadata stored in raw DNG images captured by an iPhone, against Samsung raw images that were mapped to the iPhone raw space using our Rawformer. We performed a similar comparison on rendering iPhone images with Samsung DNG metadata, both with and without our mapping.

Table \ref{ans-tab:isps} presents the results, along with those of Adobe Lightroom rendering raw/metadata taken from the same phone, which represents the best case where no mapping is needed. The results are compared against the ground-truth sRGB images of the target camera. As shown in Table \ref{ans-tab:isps}, we achieve consistent results similar to those in the referenced paper. The empirical results demonstrate that besides achieving superior performance on learnable image signal processors (ISPs), Rawformer also
exhibits a competitive edge when evaluated against commercial ISPs, like Adobe Lightroom. This indicates its robust adaptability and effectiveness across diverse ISP implementations, highlighting its potential for both research and commercial applications.

\begin{table}[!t]
\caption{\textbf{Results (PSNR/SSIM) via Lightroom (LR).} Off-diagonal results: mapping source camera raw to target camera using Rawformer before rendering with LR using target camera's metadata.}
\label{ans-tab:isps}
\centering
\scalebox{1}{
\begin{tabular}{lcc}
\multicolumn{1}{c}{} & \multicolumn{2}{c}{Target camera} \\ \cline{2-3} 
\multicolumn{1}{l|}{Source camera} & Samsung  & iPhone \\ \hline
\multicolumn{1}{l|}{Samsung} &  \textcolor{red}{22.76/0.80} & 21.98/0.77 \\
\multicolumn{1}{l|}{iPhone} & 21.42/0.76 & \textcolor{red}{22.08/0.79} \\
\bottomrule
\end{tabular}
}
\end{table}